\documentstyle[11pt,aaspp4]{article}

\def \Mpc {{\rm Mpc}}
\def \hMpc {{\rm h^{-1} Mpc}}

\def \min {{\rm min}}
\def \max {{\rm max}}
\def \lim {{\rm lim}}
\def \core {{\rm core}}
\def \fine {{\rm fine}}
\def \coarse {{\rm coarse}}
\def \model {{\rm model}}
\def \peak {{\rm peak}}

\def \data {{\rm data}}
\def \none {{\rm empty}}
\def \filled {{\rm filled}}

\def \cut {{\rm cut}}
\def \RA {{\rm RA}}
\def \DEC {{\rm DEC}}

\def \Lc {{\Lambda}}

\def \sL {{\cal L}}
\def \half {{\case{1}{2}}}

\def \BDM {\begin{displaymath}}
\def \EDM {\end{displaymath}}
\def \BEQ {\begin{equation}}
\def \EEQ {\end{equation}}
\def \BEQA {\begin{eqnarray}}
\def \EEQA {\end{eqnarray}}
\def \NN {\nonumber}
\def \BL {\begin{list}}
\def \EL {\end{list}}
\def \BENUM {\begin{enumerate}}
\def \EENUM {\end{enumerate}}
\def \BITEM {\begin{itemize}}
\def \EITEM {\end{itemize}}
\def \BARR {\begin{array}}
\def \EARR {\end{array}}

\begin{document}

\title{An Automated Cluster Finder: the Adaptive Matched Filter}
\author{Jeremy~Kepner\altaffilmark{1,2}, Xiaohui~Fan\altaffilmark{1},
Neta~Bahcall\altaffilmark{1}, James~Gunn\altaffilmark{1},
Robert~Lupton\altaffilmark{1}
and Guohong Xu\altaffilmark{1}}
\altaffiltext{1}{Princeton University Observatory, Peyton Hall, Ivy Lane,
Princeton, NJ 08544--1001 \\
(jvkepner/fan/neta/jeg/rhl/xu)@astro.princeton.edu}
\altaffiltext{2}{Current address: MIT Lincoln Laboratory, Lexington, MA}


\begin{abstract}

  We describe an automated method for detecting clusters of galaxies in
imaging and redshift galaxy surveys.  The Adaptive Matched Filter (AMF)
method utilizes galaxy positions, magnitudes, and---when
available---photometric or spectroscopic redshifts to find clusters and
determine their redshift and richness.  The AMF can be applied to most
types of galaxy surveys: from two-dimensional (2D) imaging surveys, to
multi-band imaging surveys with photometric redshifts of any accuracy
(2$\half$D), to three-dimensional (3D) redshift surveys.  The AMF
can also be utilized in the selection of clusters in cosmological
N-body simulations.  The AMF identifies clusters by finding the peaks
in a cluster likelihood map generated by convolving a galaxy survey
with a filter based on a model of the cluster and field galaxy
distributions. In tests on simulated 2D and 2$\half$D data with a
magnitude limit of $r' \approx 23.5$, clusters are detected with an
accuracy of $\Delta z \approx 0.02$ in redshift and $\sim$10\% in
richness to $z \lesssim 0.5$.  Detecting clusters at higher redshifts
is possible with deeper surveys.  In this paper we present the theory
behind the AMF and describe test results on synthetic galaxy catalogs.

\end{abstract}


\section{Introduction}

  Clusters of galaxies---the most massive virialized systems
known---provide powerful tools in the study of cosmology: from tracing
the large-scale structure of the universe (\cite{Bahcall88},
\cite{Huchra90}, \cite{Postman92}, \cite{Dalton94}, \cite{Peacock94} and
references therein) to determining the amount of dark matter on Mpc
scales (\cite{Zwicky57}, \cite{Tyson90}, \cite{Kaiser93},
\cite{Peebles93}, \cite{Bahcall95}, \cite{Carlberg96}) to studying the
evolution of cluster abundance and its cosmological implications
(\cite{Evrard89}, \cite{Peebles89}, \cite{Henry92}, \cite{Eke96},
\cite{Bahcall97}, \cite{Carlberg97}, \cite{Oukbir97}).  The above
studies place some of the strongest constraints yet on cosmological
parameters, including the mass-density parameter of the universe, the
amplitude of mass fluctuations at a scale of 8~$\hMpc$ and the baryon
fraction. 

  The availability of complete and accurate cluster catalogs needed for
such studies is limited.  One of the most used catalogs, the Abell
catalog of rich clusters (\cite{Abell58}, and its southern counterpart
\cite{Abell89}), has been extremely useful over the past four decades. 
This catalog, which contains $\sim$4000 rich clusters to $z \lesssim
0.2$ over the entire high latitude sky, with estimated redshifts and
richnesses for all clusters, was constructed by visual selection from
the Palomar Sky Survey plates, using well-defined selection criteria. 
The Zwicky cluster catalog (\cite{Zwicky61}) was similarly constructed
by visual inspection.

  The need for new, objective, and accurate large-area cluster
catalogs to various depths is growing, following the important use of
clusters in cosmology.  Large area sky surveys using CCD imaging in one or
several colors, as well as redshift surveys, are currently planned or
underway, including, among others, the Sloan Digital Sky Survey (SDSS).
Such surveys will provide the data needed for constructing accurate
cluster catalogs that will be selected in an objective and automated
manner. In order to identify clusters in the new galaxy surveys, a
robust and automated cluster selection algorithm is needed.  We propose
such a method here.

  Cluster identification algorithms have typically been targeted at
specific surveys, and new algorithms have been created as each survey is
completed.  \cite{Abell58} was the first to develop a well-defined
method for cluster selection, even though the identification was carried
out by visual inspection (see, e.g., \cite{McGill90} for a analysis of
this method).  Other algorithms have been created for the APM survey
(\cite{Dalton94}, \cite{Dalton97}; see \cite{Schuecker98} for a variant
of this method), the Edinburgh-Durham survey (ED; \cite{Lumsden92}), and
the Palomar Distant Cluster Survey (PDCS; \cite{Postman96}; see also
\cite{Kawasaki97} for a variant of this method; and \cite{Kleyna97} for
an application this method to finding dwarf spheroidals).  All the above
methods were designed for and applied to two-dimensional imaging
surveys. 

  In this paper we present a well defined, quantitative method, based
on a matched filter technique that expands on some of the previous
methods and provides a general algorithm that can be used to identify
clusters in any type of survey.  It can be applied to 2D imaging
surveys, 2$\half$D surveys (multi-band imaging with photometric
redshift estimates of any accuracy), 3D redshift surveys, as well as
combinations of the above (i.e. some galaxies with photometric
redshifts and some with spectral redshifts).  In addition, this
Adaptive Matched Filter (AMF) method can be applied to identify
clusters in cosmological simulations.

  The AMF identifies clusters by finding the peaks in a cluster
likelihood map generated by convolving a galaxy survey (2D, 2$\half$D or
3D) with a filter which models the cluster and field galaxy
distribution. The peaks in the likelihood map correspond to locations
where the match between the survey and the filter is maximized.  In
addition, the location and value of each peak also gives the best fit
redshift and richness for each cluster.  The filter is composed of
several sub-filters that select different components of the survey: a
surface density profile acting on the position data, a luminosity profile
acting on the apparent magnitudes, and, in the 2$\half$D and 3D cases, a
redshift cut acting on the estimated redshifts.

  The AMF is adaptive in three ways. First, the AMF adapts to the errors
in the observed redshifts (from no redshift information (2D), to
approximate (2$\half$D) or measured redshifts (3D)).  Second, the AMF
uses the location of the galaxies as a ``naturally'' adaptive grid to
ensure sufficient spatial resolution at even the highest redshifts. 
Third, the AMF uses a two step approach that first applies a coarse
filter to find the clusters and then a fine filter to provide more
precise estimates of the redshift and richness of each cluster.

  We describe the theory of the AMF in \S2 and its implementation in
\S3.  We generate a synthetic galaxy catalog to test the AMF in \S4 and
present the results in \S5.  We summarize our conclusions in \S6.

\section{Derivation of the Adaptive Matched Filter}

  The idea behind the AMF is the matching of the data with a filter
based on a model of the distribution of galaxies. The model describes
the distribution in surface density, apparent magnitude, and redshift of
cluster and field galaxies.  Convolving the data with the filter
produces a cluster probability map whose peaks correspond to the
location of the clusters. Here we describe the theory behind the AMF. 
We present the underlying model in \S2.1, the concept of the cluster
overdensity in \S2.2, two likelihood functions derived under different
assumptions about the galaxy distribution in \S2.3, and the extension of
the likelihood functions to include estimated redshifts in \S2.4.

\subsection{Cluster and Field Model}
  The foundation of the AMF is the model of the total number density of
galaxies per solid angle ($d\Omega = 2 \pi \theta d\theta$) per apparent
luminosity ($dl$) around a cluster at redshift $z_c$ with richness $\Lc$
  \BEQ
       n_\model(\theta,l;z_c) d\Omega dl = [n_f(l) + \Lc n_c(\theta,l;z_c)] d\Omega dl ~ ,
  \EEQ
where $\theta$ is the angular distance from the center of the cluster,
and $n_f$ and $\Lc n_c$ are the number densities due to the field and
the cluster. The field number density is taken directly from the global
``number--magnitude'' per square degree relation.  The cluster number
density is the product of a projected density profile (see Appendix A)
and a luminosity profile, both shifted to the redshift of the cluster
and transformed from physical radius and absolute luminosity to angular
radius and apparent luminosity (i.e. flux)
  \BEQ
       \Lc n_c(\theta,l;z_c) = \Lc \Sigma_c(r) \left ( \frac{dr}{d\theta} \right )^2
                               \phi_c(L) \left ( \frac{dL}{dl} \right )
  \EEQ
where the projected radius $r$ and absolute luminosity $L$ at the cluster
redshift are given by
  \BEQA
        r(\theta,z_c) &=& \frac{\theta d_c}{1 + z_c} ~ , \NN \\
        L(l,z_c,\mu)  &=& 4 \pi D^2(z_c,\mu) l ~ .
  \EEQA
Here $d_c$ is the angular size distance corresponding to $z_c$, and
$D$ is the luminosity distance (with K correction) of a galaxy of
spectral type $\mu$ (e.g. elliptical, spiral or irregular) (see
Appendix B). The factor of $1 + z_c$ in the radius relation converts
from comoving to physical units (the cluster profile can be defined with
either comoving or physical units). For the cluster luminosity profile
a Schechter function is adopted
  \BEQ
        \phi_c(L) dL \propto (L/L^*)^{-\alpha} e^{-L/L^*} d(L/L^*),
  \EEQ 
where $\alpha \simeq 1.1$ (\cite{Schechter76}, \cite{Binggeli88},
\cite{Loveday92}).

  The model is completed by choosing a normalization for the
radial profile $\Sigma_c$ and the luminosity profile $\phi_c$. The
choice of normalization is arbitrary, but has the effect of determining
the units of the richness $\Lc$.  We choose to normalize so that the
total luminosity of the cluster is equal to
  \BEQA
      L_c( < r_\max) & = & \Lc L^* \NN \\
                   & = & \Lc \int L(l,z_c) n_c(\theta,l;z_c) d\Omega dl \NN \\
                   & = & \Lc \int_0^{r_\max} \Sigma_c(r) 2 \pi r dr
                         \int_0^\infty \phi_c(L) L dL ~ .
  \EEQA
The richness parameter $\Lc$ thus describes the total cluster
luminosity (in units of $L^*$) within $r_\max$.  For $r_\max =
1~\hMpc$, the richness $\Lc( \leq 1~\hMpc)$ relates to Abell's richness
$N_A$ (within 1.5 $\hMpc$) as $N_A \approx \frac{2}{3} \Lc$. For example,
Abell richness class $\geq 1$ clusters ($N_A \geq 50$) correspond to $\Lc \geq
75$ (\cite{Bahcall93}). Multiplying by $L$ in equation (5) allows $\phi_c$ to be
integrated down to zero luminosity, thus insuring that the total
luminosity is equal to $\Lc L^*$ regardless of the the apparent
luminosity limit of the survey $l_\min$.  The above constraint can be
implemented by simply multiplying $\Sigma_c$ and $\phi_c$ by any
appropriate constants (e.g., normalizing the first integral to one and
setting the second integral equal to $L^*$).

\subsection{Cluster Overdensity}
  Clusters, by definition, are density enhancements above the field. 
To quantify this we introduce the cluster overdensity $\Lc \Delta$,
which is the sum of the individual overdensities of the galaxies $\Lc
\delta_i$.  $\Delta$ and $\delta_i$ are defined subsequently.

  For a given cluster location on the sky let $\theta_i$ and $l_i$ be
the angular separation from the cluster center and the apparent
luminosity of the $i$th galaxy, respectively. For a specific cluster
redshift $z_c$ we only need consider galaxies inside the maximum
selected cluster radius (Appendix A)
  \BEQ
        \theta_\max(z_c) = \frac{(1 + z_c) r_\max}{d_c} ~.
  \EEQ
The apparent local overdensity at the location of the $i$th galaxy is
given by $\Lc \delta_i$, where
  \BEQ
        \delta_i = \frac{n_c^i}{n_f^i}
                 = \frac{n_c(\theta_i,l_i;z_c) d\Omega dl }
                        {n_f(l_i) d\Omega dl} ~ .
  \EEQ
The apparent overdensity of the cluster as measured from the data is
  \BEQ
        \Lc \Delta_\data = \Lc \sum_i \delta_i ~ ,
  \EEQ
where the sum is carried out over all galaxies within $\theta_\max$. 
The cluster overdensity can also be calculated from the model
  \BEQ
        \Lc \Delta_\model =
        \Lc \int^\infty_{l_\min} \int^{\theta_\max}_0
            \delta(\theta,l;z_c) \Lc n_c(\theta,l;z_c) d\Omega dl ~ .
  \EEQ
Equating $\Delta_\data$ with $\Delta_\model$ it is possible to solve for $\Lc$
  \BEQ
        \Lc = \frac{\sum_i \delta_i}{\int \delta ~ n_c} ~ .
  \EEQ
The term in the denominator is simply the model cluster overdensity when
$\Lc = 1$. As we shall see in the subsequent sections, the positions of
clusters, both in angle and in redshift, correspond to the locations of
maxima in the cluster overdensity.

\subsection{Likelihood Functions}
  Having defined the model we now discuss how to find clusters in a
galaxy catalog and determine the best fit redshift and richness $z_c$
and $\Lc$ for each cluster. The basic scheme is to define a function
$\sL$, which is the likelihood that a cluster exists, and to find the
parameters that maximize $\sL$ at a given position over the range of
possible values of $z_c$ and $\Lc$.  This procedure is carried out over
the entire survey area and generates a likelihood map. The clusters are
found by locating the peaks in the likelihood map. The map grid can be
chosen by various means, such as a uniform grid or by using the galaxy
positions themselves.

  A variety of likelihood functions can be derived, depending on the
assumptions that are made about the distribution of the galaxies.  The
AMF uses two likelihood functions whose derivations are given in
Appendix C. A cluster is identified, and its redshift and richness,
$z_c$ and $\Lc$, are found by maximizing $\sL$.  Typically this is
accomplished by first taking the derivative of $\sL$ with respect to
$\Lc$ and setting the result equal to zero
  \BEQ
       \frac{\partial \sL}{\partial \Lc} = 0 ~ .
  \EEQ
One can compute $\Lc$ from the above equation either directly or
numerically.  The resulting value of $\Lc$ can be inserted back into
the expression for $\sL$ to obtain a value for $\sL$ at the specified
redshift.  Repeating this procedure for different values of $z_c$ it is
possible to find the maximum likelihood and the associated best value
of the cluster redshift, as well as the best cluster richness at this
redshift.

  The first likelihood, which we call the coarse likelihood, assumes
that if we bin the galaxy catalog, then there will be enough galaxies
in each bin that the distribution can be approximated as a Gaussian.
This assumption, while not accurate, provides a coarse
likelihood function $\sL_\coarse$ which is linear, easy to compute and
corresponds to the apparent overdensity
  \BEQ
        \sL_\coarse = \Lc_\coarse \sum_i \delta_i ~~,
  \EEQ
where
  \BEQ
        \Lambda_\coarse = \frac{
              \sum_i \delta_i
           }{
              \int \delta(\theta,l) n_c(\theta,l) d\Omega dl
          }
  \EEQ
(see Appendix C).

  The second likelihood, which we call the fine likelihood, assumes that
the galaxy count in a bin can be modeled with a Poisson distribution---an
assumption which is nearly always correct. The resulting
likelihood function $\sL_\fine$ is non-linear and requires more
computations to evaluate, but should provide a more accurate estimate of
the cluster redshift and richness.  The formula for $\sL_\fine$ is
  \BEQ
        \sL_\fine = - \Lambda_\fine  N_c + \sum_i \ln[1 + \Lambda_\fine \delta_i] ~ ,
  \EEQ
where $\Lc_\fine$ is computed by solving
  \BEQ
       N_c = \sum_i \frac{\delta_i}{1 + \Lambda_\fine \delta_i} ~ ,
  \EEQ
and $N_c$ is the number of galaxies one would expect to see in a cluster with
$\Lc = 1$ (see Appendix C).

\subsection{Including Estimated Redshifts}
  So far we have considered cluster selection in purely 2D imaging
surveys with no estimated redshift information. The inclusion of
estimated galaxy redshifts, as in 2$\half$D or 3D surveys, should
improve the accuracy of the resulting cluster redshifts.  In addition,
estimated redshifts can separate background galaxies (noise)  from
cluster galaxies (signal) more easily, allowing the detection of
considerably poorer clusters than if estimated redshifts were not used.

  Galaxy redshift information can be obtained, for example, from
multi-color photometric surveys (via photometric redshifts estimates) or
from direct spectroscopic determination of galaxy redshifts.  We now
discuss how to extend the AMF described above to include such redshifts. 
The galaxy redshifts that we use {\it within a given survey} can range
from very precise measured redshifts, to only approximate photometric
redshifts (with varied accuracy), to no redshift information---all in
the same analysis (i.e.  adapting to different redshift information for
each galaxy in the survey).  The available redshifts provide essentially
a third filter, in addition to the spatial and luminosity filters used
in the 2D case.  In practice, we use the estimated redshift information
of each galaxy to narrow the window of the AMF search following the same
basic method described above. 

  One of the benefits of laying down the theoretical framework for the
AMF is the easy means by which estimated redshifts can now be included. 
Let $z_i$ and $\sigma_z^i$ be the estimated redshift and redshift error
of the $i$th galaxy, and let the factor $w$ determine how wide a region
we select around each $z_c$ for the likelihood analysis.  Inclusion of
the additional redshift information is accomplished by simply limiting
the sums in $\sL_\coarse$ and $\sL_\fine$ to those galaxies that also
satisfy $|z_i - z_c| < w \sigma_z^i$.  Note that this procedure allows
the usage of redshift information with variable accuracy in the same
survey---i.e., some galaxies with measured redshifts, some with
estimated redshifts and some with no redshifts at all. 

  The richness $\Lambda_\coarse$ is dominated by the cluster galaxies;
as long as $w$ is sufficiently large (e.g. $w \approx 3$) then the
resulting richnesses will be unbiased.  However, if the redshift errors
are large (i.e., a large fraction of the depth of the galaxy catalog), it
may be desirable to use a smaller value of $w \approx 1$.  In this case
a small correction needs to be applied to the richness to account for
the small fraction of cluster galaxies that are eliminated by the
redshift cut. If the redshift errors are Gaussian, the desired
correction can be obtained by multiplying the predicted cluster
overdensity by the standard error function
  \BEQ
        \sqrt{\frac{2}{\pi}}
                          \int^w_0 e^{-\frac{1}{2} t^2} dt ~ .
  \EEQ

  A similar issue arises with $\Lambda_\fine$ when the field galaxies
are eliminated by using estimated redshifts. To obtain the correct
value of $\Lambda_\fine$ requires modeling the contribution of the
eliminated field galaxies to equation (15), which can be done either
analytically or numerically via Monte Carlo methods.

\section{Implementation}
\begin{figure}
\plotone{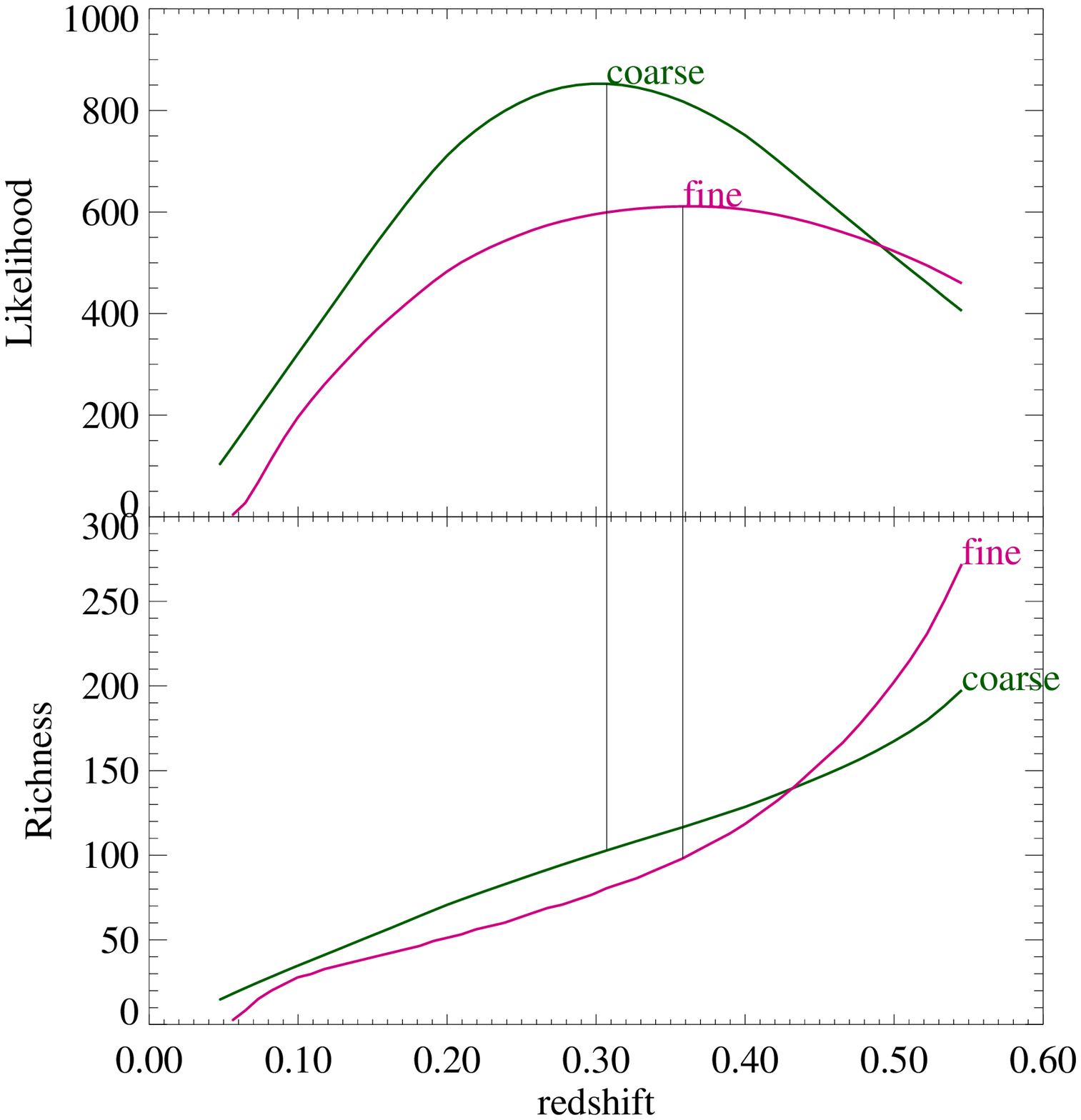}
\caption{
\label{fig:likelihood:a}
  Plots of the ``2D'' (i.e., $\sigma_z \rightarrow \infty$) likelihood
and richness as a function of redshift as computed from the coarse and
fine matched filters.  The input cluster has $z_c = 0.35$ and a
richness $\Lc = 100$ (corresponding approximately to Abell richness
class 1).
} \end{figure}

\begin{figure}
\plotone{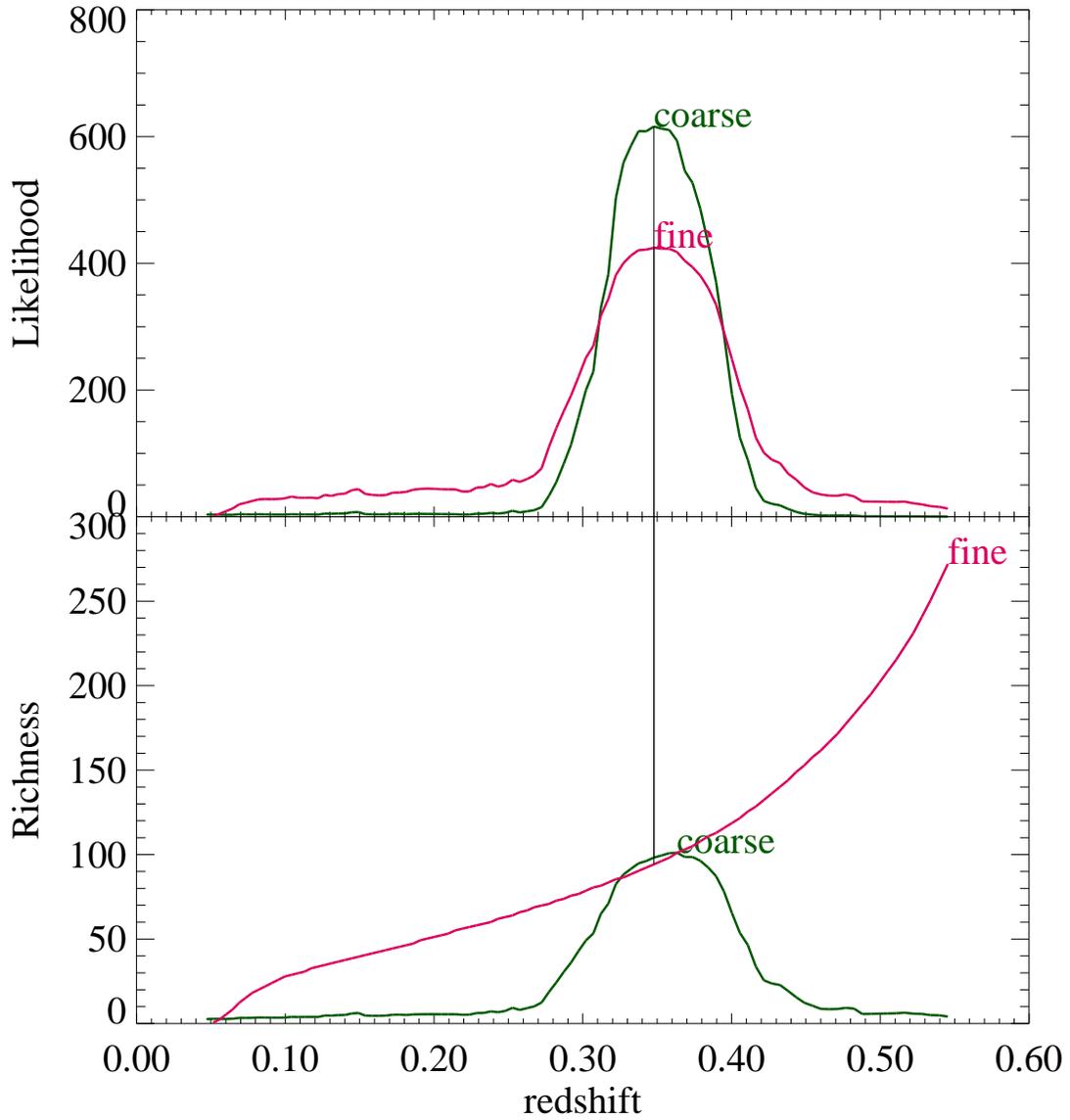}
\caption{
\label{fig:likelihood:b}
  Plots of the ``2$\half$D'' ($\sigma_z \sim 0.05$) likelihood and
richness as a function of redshift as computed from the coarse and fine
matched filters.  The input cluster has $z_c = 0.35$ and a richness
$\Lc = 100$. Although $\Lc_\fine$ is not peaked like $\Lc_\coarse$ this
difference does not diminish the accuracy since the location of $z_c$
is determined entirely by $\sL_\fine$, which is sharply peaked.
} \end{figure}

  The likelihood functions derived in the previous section represent
the core of the AMF cluster detection scheme. Both likelihood functions
begin with picking a grid of locations over the survey area for which
$\sL$ and $\Lc$ are computed over a range of redshifts. 
Figures~\ref{fig:likelihood:a} and~\ref{fig:likelihood:b} show the
functions $\sL(z)$ and $\Lc(z)$ for a position on the sky located at
the center of a $\Lc = 100$ cluster ($\sim\frac{2}{3}$ the richness or
luminosity of Coma) at $z = 0.35$ (the details of the test data are
given in the next section). Figure~\ref{fig:likelihood:a} shows the
results when no redshift estimate exists (i.e., $\sigma_z \rightarrow
\infty$), and figure~\ref{fig:likelihood:b} shows what happens when
using estimated redshifts with $0.03 < \sigma_z < 0.06$.

  Figures~\ref{fig:likelihood:a} and~\ref{fig:likelihood:b} illustrate
three of the basic properties of these likelihood functions, which were
discussed in the previous section: 1. the dramatic effect of including
estimated redshifts on sharpening the peak in $\sL$, which makes
finding clusters much easier; 2. the qualitative difference in the form
of $\Lc_\fine(z)$ and $\Lc_\coarse(z)$ in figure~\ref{fig:likelihood:b}
that arises from the fact that $\Lc_\coarse$ is a simple function of
$\sL_\coarse$ while $\Lc_\fine$ is a complicated function of
$\sL_\fine$; 3. the shortward bias in the cluster redshift as
computed from $\sL_\coarse$ in 2D (figure~\ref{fig:likelihood:a}), which is a
general bias in the 2D coarse likelihood function (for a discussion of
how to correct for this bias see \cite{Postman96}).   In the 3D case,
the coarse likelihood of rich clusters has comparable accuracy to the
fine likelihood. For poorer clusters ($\Lc < 50$), the coarse
likelihood yields higher richness estimates than the true values; this
is a result of the Gaussian assumption used for deriving the coarse
likelihood which worsens for poorer clusters.

  Implementation of the AMF cluster selection consists of five steps:
(1) reading the galaxy catalog, (2) defining and verifying the model,
(3) computing $\sL_\coarse$ over the entire galaxy survey over a range
of redshifts, (4) finding clusters by identifying peaks in the
$\sL_\coarse$ map, and (5) evaluating $\sL_\fine$ and obtaining a more
precise determination of each cluster's redshift and richness.

\subsection{Reading the Galaxy Catalog}
  The first step of the AMF is reading the galaxy catalog over a
specified region of the sky. The galaxy catalog consists of four to six
quantities for each galaxy, $i$: the position on the sky RA$_i$ and
DEC$_i$, the apparent luminosity (i.e. flux) in specific band $l_i$,
the Hubble type used for determining the K correction $\mu_i$ (e.g., E,
Sa or Sc), the estimated redshift $z_i$ and estimated error
$\sigma_z^i$.  In the case of a single band survey, where no
photometric redshifts are possible, $z_i$ and $\sigma_z^i$ will not
exist.  In addition to these local quantities, the following global
quantities are computed from the catalog: the apparent luminosity limit
$l_\min$, the mean estimated error $\bar{\sigma}_z$.  Finally, it is
necessary to set the minimum and maximum cluster redshifts for the
cluster search $z_\min$ and $z_\max$.

\subsection{Model Definition and Verification}
  The primary model components which need to be specified are: the
cluster radial profile $\Sigma_c(r)$, the cluster luminosity profile
$\phi_c(L)$, a normalization convention for the cluster that sets the
units of richness $\Lc$, and the field number density versus apparent
luminosity distribution $n_f(l)$. In addition, a specific cosmological
model needs to be chosen along with K corrections for each Hubble type,
from which the angular distance $d(z)$ and luminosity distance
$D(z,\mu)$ can be computed.  The observed galaxy catalog needs to be
tested for its consistency with the model parameters (such as $n_f(l)$). 
While this is unnecessary for the simulated galaxy catalog used below
(where the model used to find clusters is the same as the one used to
generate the catalog), we show some of the consistency checks for
illustration purposes (see Appendix D).

\subsection{$\sL_\coarse$ Mapping}
  Mapping out the coarse likelihood function begins with picking a grid
covering the survey area.  The most straightforward choice is a uniform
grid over the area covered, which is conceptually simple and makes
finding peaks in the likelihood easy.  However, a uniform grid needs to
be exceedingly fine in order to ensure adequate resolution at the
highest redshifts, and leads to unacceptable computer memory requirements.
Another choice for the grid locations is to use the positions of the
galaxies themselves.  The galaxy positions are a ``naturally'' adaptive
grid that guarantees sufficient resolution at any redshift while also
eliminating unnecessary points in sparse regions. For this reason we
choose to use the galaxy positions as the grid locations.

  At each grid location (i.e. each galaxy position) $\sL_\coarse$ is
evaluated over a range of redshifts from $z_\min$ to $z_\max$.   The
number of redshift points is set so there is adequate coverage for the
given value of $\bar{\sigma}_z$.  Finding the maximum of the likelihood
sets the values of the likelihood, redshift and richness at this point:
$\sL_\coarse^i$, $z_\coarse^i$, and $\Lc_\coarse^i$.  When this process
has been completed for all the galaxies, the result is an irregularly
gridded map $\sL_\coarse^i(\RA_i,\DEC_i,z_\coarse^i)$.  The peaks in
this map correspond to the locations of the clusters.

\subsection{$\sL_\coarse$ Cluster Selection}
  Finding peaks in 3D regularly gridded data is straightforward.
Finding the peaks in the irregularly gridded function
$\sL_\coarse^i(\RA_i,\DEC_i,z_\coarse^i)$ is more
difficult.  There are several possible approaches. We present a simple
method which is sufficient for selecting rich clusters.  More sophisticated
methods will be necessary in order to find poor clusters that are close
to rich clusters. 

\begin{figure}
\plotone{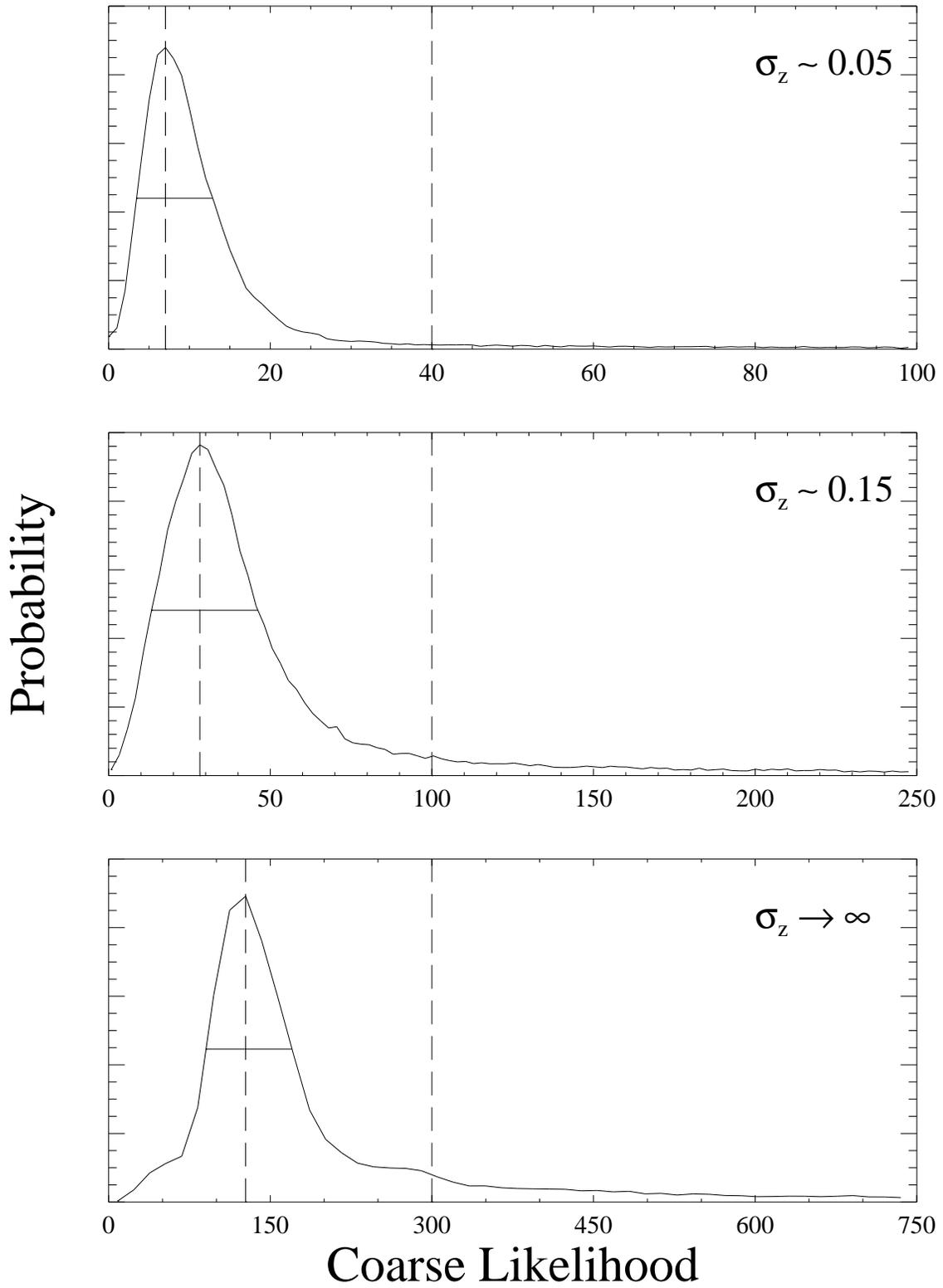}
\caption{
\label{fig:outputclusters:h}
  Distribution of $\sL_\coarse$ values.  The range has been chosen so
the distributions appear similar. The left dashed line denotes
the value of the peak of the distribution.  The right dashed line shows
the value of $\sL_\cut$ that was used.  The horizontal solid line shows
the FWHM around the peak.  The significance levels of $\sL_\cut$ from
top to bottom are approximately 8-$\sigma$, 5-$\sigma$ and 5-$\sigma$.
} \end{figure}

  As a first step we eliminate all low likelihood points $\sL_\coarse^i <
\sL_\cut$, where $\sL_\cut$ is the nominal detection limit, which is
independent of richness or redshift.  $\sL_\cut$
can be estimated from the distribution of the $\sL_\coarse^i$
values (figure~\ref{fig:outputclusters:h}).  The peak in the
distribution corresponds to the background while the long tail
corresponds to the clusters.  Assuming the background is a Gaussian whose
mean can be estimated from the location of the peak and whose standard
deviation is 0.43 times the Full-Width-Half-Maximum (FWHM), then a given
value of $\sL_\cut$ will lie $N_\sigma$ standard deviations from the peak
\BEQ
    N_\sigma \sim \frac{\sL_\cut - \sL_\peak}{0.43 ~ {\rm FWHM}} ~ .
\EEQ
The values of $\sL_\cut$ shown in figure~\ref{fig:outputclusters:h}
were chosen so that $N_\sigma \sim 5$, which was sufficiently high that
no false detections occurred. To first order, $\sL_\cut \propto
\bar{\sigma}_z/\bar{z}$, where $\bar{z}$ is the average depth of the
survey and is a function of $l_\min$.

  Step two consists of finding the largest value of $\sL_\coarse^i$,
which is by definition the first and most overdense cluster.  The third step
is to eliminate all points that are within a certain radius and
redshift of the cluster.  Repeating steps two and three until there are
no points left results in a complete cluster identification (above $\sL_\cut$),
with a position, redshift and richness ($\propto$ total luminosity) estimate
for each cluster.


\subsection{$\sL_\fine$ Evaluation}
  The angular position, redshift and richness of the clusters
determined by the $\sL_\coarse$ selection are adequate, but can be
complemented by determining the redshift and richness from $\sL_\fine$.
Recall that $\sL_\fine$ requires 10 to 100 times as many operations to
evaluate as compared with $\sL_\coarse$, and applying it to every
single galaxy position can be prohibitive.  Evaluating $\sL_\fine$ on
just the clusters found with $\sL_\coarse$ is trivial and worth doing
as it should provide more accurate estimates of the cluster redshift
and richness because of the better underlying assumptions that went
into its derivation.  Thus, the final step in our AMF implementation is
to compute $z_\fine^i$ and $\Lc_\fine^i$ using $\sL_\fine$ on each of
the cluster positions obtained with $\sL_\coarse$.

\subsection{Implementation Summary}
  We summarize the above implementation in the following step-by-step list.
\BENUM
  \item Read galaxy catalog
  \BITEM
    \item Read in RA$_i$, DEC$_i$, $l_i$, $z_i$, $\sigma_z^i$ and
          $\mu_i$ for each galaxy.
    \item Pick $l_\min$ (survey limit), $z_\min$ and $z_\max$ (cluster search limits),
          and compute average distance error $\bar{\sigma}_z$.
  \EITEM
  \item Model definition
  \BITEM
    \item Choose $\Sigma_c$, $\phi_c$, $n_f$, $d(z_c)$ and $D(z,\mu)$.
    \item Normalize $\Sigma_c$ and $\phi_c$.
    \item Verify galaxy distributions in $l$ and $z$ with those predicted by model.
  \EITEM
  \item $\sL_\coarse$ mapping
  \BITEM
    \item For each galaxy location (RA$_i$, DEC$_i$) choose a range
          of redshifts within $z_\min$ and $z_\max$ with a step size
          no larger that one half $\bar{\sigma}_z$.
    \item Compute $\sL_\coarse$ (eq. 12) and $\Lc_\coarse$ (eq. 13) for each redshift.
          Set $\sL_\coarse^i$, $\Lc_\coarse^i$ and $z_\coarse^i$ equal to the
          values at the maximum of $\sL_\coarse$.
  \EITEM
  \item Find peaks in the $\sL_\coarse$ map
  \BITEM
    \item Compute $\sL_\cut$ from the $\sL_\coarse^i$ distribution (e.g. 5-$\sigma$ cut).
    \item Find all local maxima in $\sL_\coarse^i(\RA_i,\DEC_i,z_\coarse^i)$
          where $\sL_\coarse^i > \sL_\cut$; these are the clusters.
  \EITEM
  \item Refine cluster redshift and richness estimates with $\sL_\fine$
  \BITEM
    \item At the RA and DEC of each cluster found with $\sL_\coarse$
          evaluate $\sL_\fine$ over the same range of redshifts within
          $z_\min$ and $z_\max$.
    \item Compute $\sL_\fine$ (eq. 14) and $\Lc_\fine$ (eq. 15) for each redshift.
          Set $\sL_\fine^i$, $\Lc_\fine^i$ and $z_\fine^i$ equal to the
          values at the maximum of $\sL_\fine$.  These provide the best
          estimates for the cluster richness and redshift.
   \EITEM
\EENUM

  In the next section we describe a simulated galaxy catalog used to
test the above AMF implementation.

\section{Simulated Test Data}
  Our test data consists of 72 simulated clusters with different
richnesses and redshifts placed in a simulated field of randomly
distributed galaxies (for a non-random distribution of field galaxies
see section 5).  The clusters range from groups to rich clusters with
total luminosities from 10 $L^*$ to 300 $L^*$ (corresponding to Abell
richness counts of approximately 7 to 200, or richness classes $\ll$0 to
$\sim$4) and are distributed in redshift from 0.1 to 0.5.  The
luminosity profile is a Schechter function with $\alpha = 1.1$.  The
radial profile is a Plummer law given in Appendix A with $r_\max =
1~\hMpc$ and $r_\core = 0.1~r_\max$.  The number and absolute luminosity
of the field galaxies were generated from a Schechter function, using
the field normalization $\phi^* = 1.08 \times 10^{-2}~{\rm
h}^3\Mpc^{-3}$ (\cite{Loveday92}). 

  Three different Hubble types were used, E, Sa and Sc, with K
corrections taken from \cite{Poggianti97}.  Each galaxy in a cluster was
randomly assigned a Hubble type so that 60\% were E, 30\% were Sa, and
10\% were Sc; each galaxy in the field is randomly assigned a Hubble
type so that 40\% were E, 30\% were Sa, and 30\% were Sc. Knowing the
redshift of each galaxy, its absolute luminosity and Hubble type, we can
compute the apparent luminosity.  All galaxies with apparent luminosity
below $r' \approx 23.5$ (the anticipated limit of the SDSS) were
eliminated.  This limit resulted in a field number density of
$\sim$5000 galaxies per square degree.

\begin{figure}
\plotone{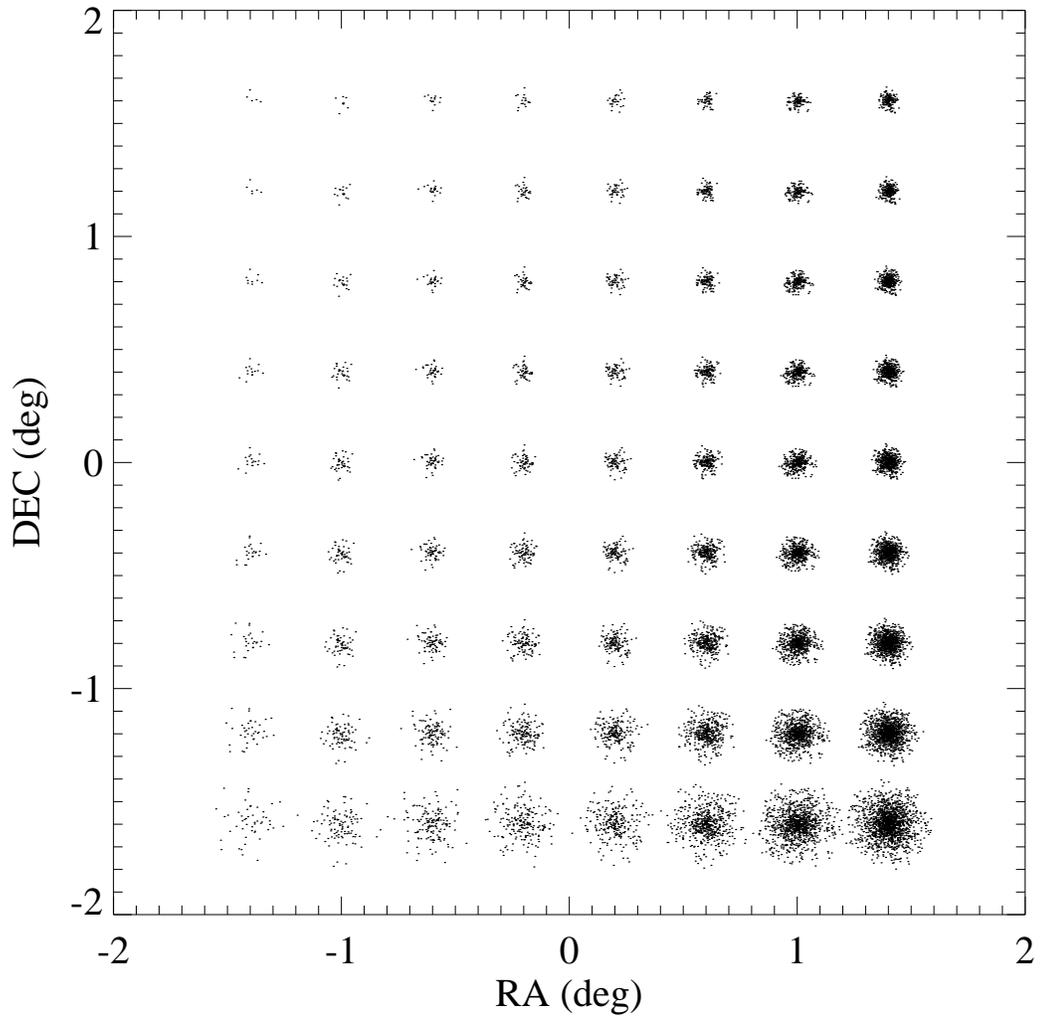}
\caption{
\label{fig:inputclusters:a}
  Angular positions of all the simulated cluster galaxies with no
background galaxies.
} \end{figure}

\begin{figure}
\plotone{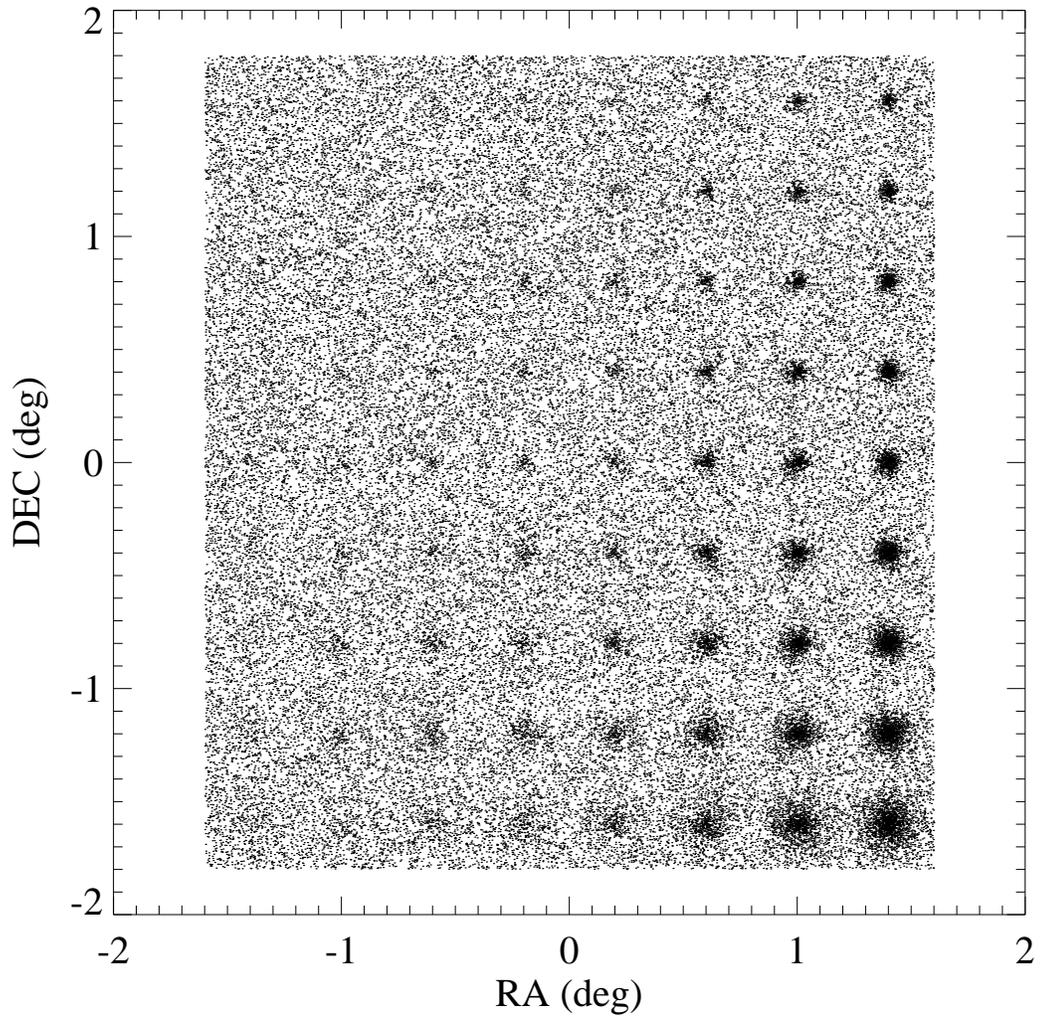}
\caption{
\label{fig:inputclusters:b}
  Angular positions of all the simulated cluster and field galaxies.
} \end{figure}

\begin{figure}
\plotone{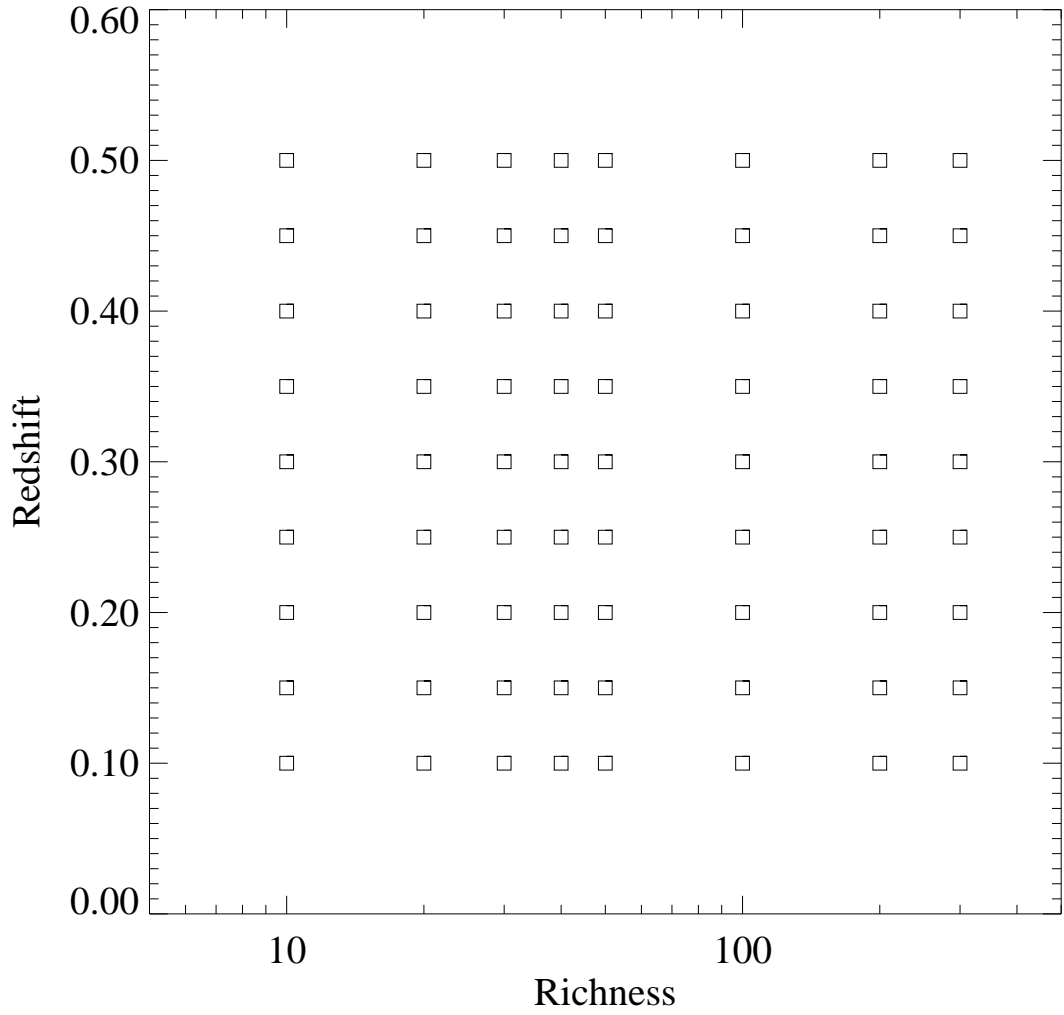}
\caption{
\label{fig:inputclusters:c}
  Richness and redshift of each of the simulated clusters.  In this
view the cluster parameters match the same 8 by 9 grid used in
Figure~\ref{fig:inputclusters:a}.
} \end{figure}

  To facilitate the subsequent analysis and interpretation of the
results, the clusters were placed on an 8 by 9 grid. The cluster
centers were separated by 0.4 degrees, which results in the test data
having dimensions of 3.2 by 3.6 degrees.  The distribution of the
cluster galaxies in RA and DEC is shown in
figure~\ref{fig:inputclusters:a}, where each column has the same
richness while each row of clusters is at the same redshift (see
figure~\ref{fig:inputclusters:c}).  From left to right the richnesses
are $\Lambda =$ 10, 20, 30, 40, 50, 100, 200, and 300.  From bottom to
top the redshifts are $z_c =$ 0.1, 0.15, 0.2, 0.25, 0.3, 0.35, 0.4,
0.45, and 0.5. In all, the clusters contained some 30,000 galaxies,
over half of which lie in the richness 200 and 300 clusters. The
randomly generated field contained approximately 50,000 galaxies. The
distribution of all the galaxies in RA and DEC is shown in
figure~\ref{fig:inputclusters:b}.

  So far, the redshifts for the galaxies are exact.  If we assume that
redshift errors are Gaussian, then we can easily generate offsets to
the true positions if we know the standard deviation of the
distribution $\sigma_z$.  Not all the galaxies will have the same
estimated redshift error $\sigma_z^i$, and, in the case of photometric
redshifts, these values can be expected to vary by about a factor of
two (\cite{Connolly95}; \cite{Yee98}).  We model this effect by first randomly
generating the estimated redshift errors from a uniform distribution
over a specified range (e.g., $0.03 < \sigma_z^i < 0.06$).  The
offsets from the true redshifts are then randomly generated from
Gaussian distributions with standard deviations corresponding to
$\sigma_z^i$ values.  The estimated redshifts are computed by adding
the offsets to the true redshift.  Thus, a dataset with $\bar{\sigma}_z
\sim 0.05$ refers to $0.03 < \sigma_z^i < 0.06$.

\section{Results and Discussion}
\begin{figure}
\plotone{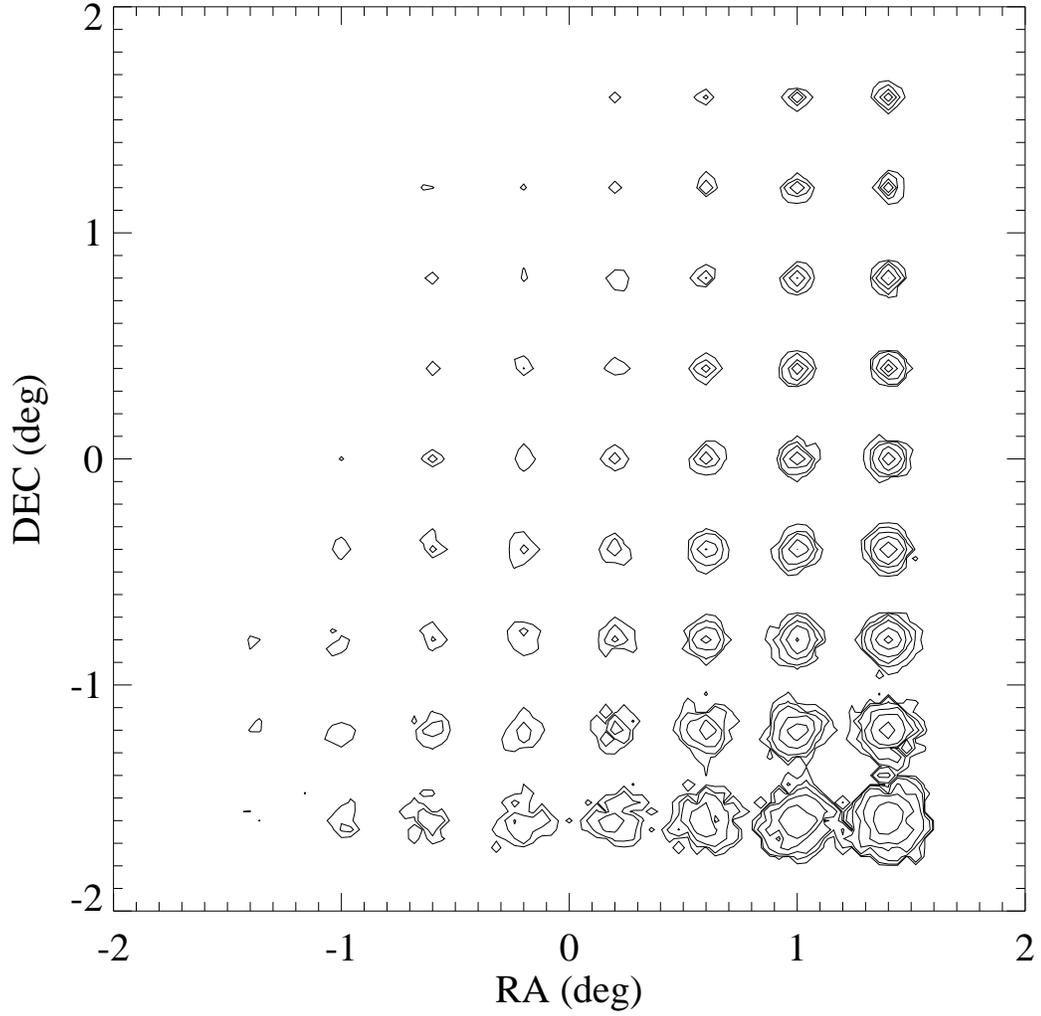}
\caption{
\label{fig:outputclusters:i}
  Contour plot of $\sL_\coarse$ computed from simulated data with photometric
redshift errors $\sigma_z \sim 0.05$. Contour levels begin at
$\sL_\coarse = 40$ and increase by a factor of 3 with each subsequent
level.
} \end{figure}

\begin{figure}
\plotone{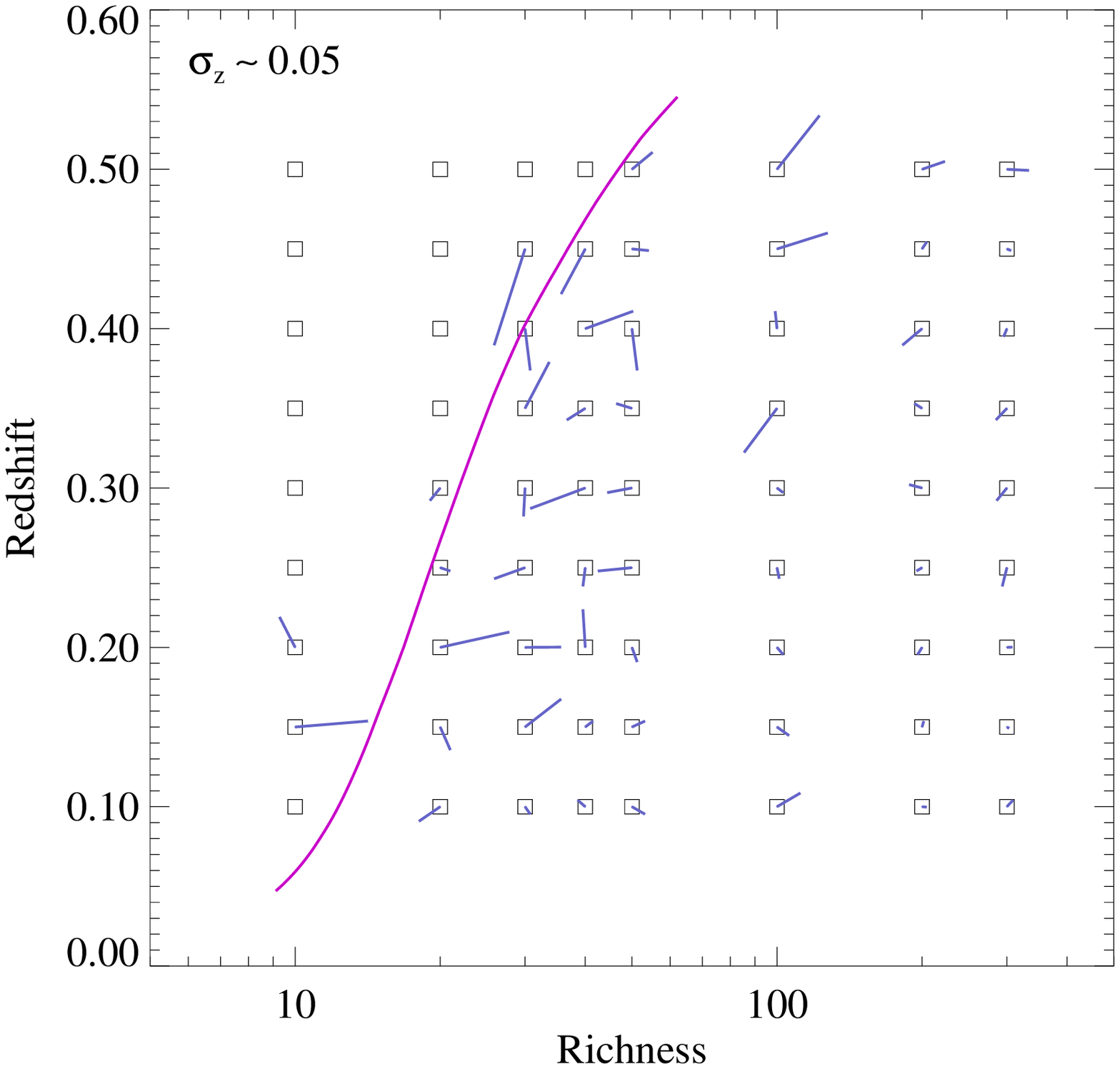}
\caption{
\label{fig:outputclusters:b}
  Richness and redshift of each of the input clusters (boxes) with the
short lines indicating the corresponding values determined from the AMF
fine filter. The long curved line indicates the approximate detection
limit and is computed by inserting $\sL_\cut = 40$ into equation (12) and
solving for $\Lc$ as a function $z_c$.
} \end{figure}

\begin{figure}
\plotone{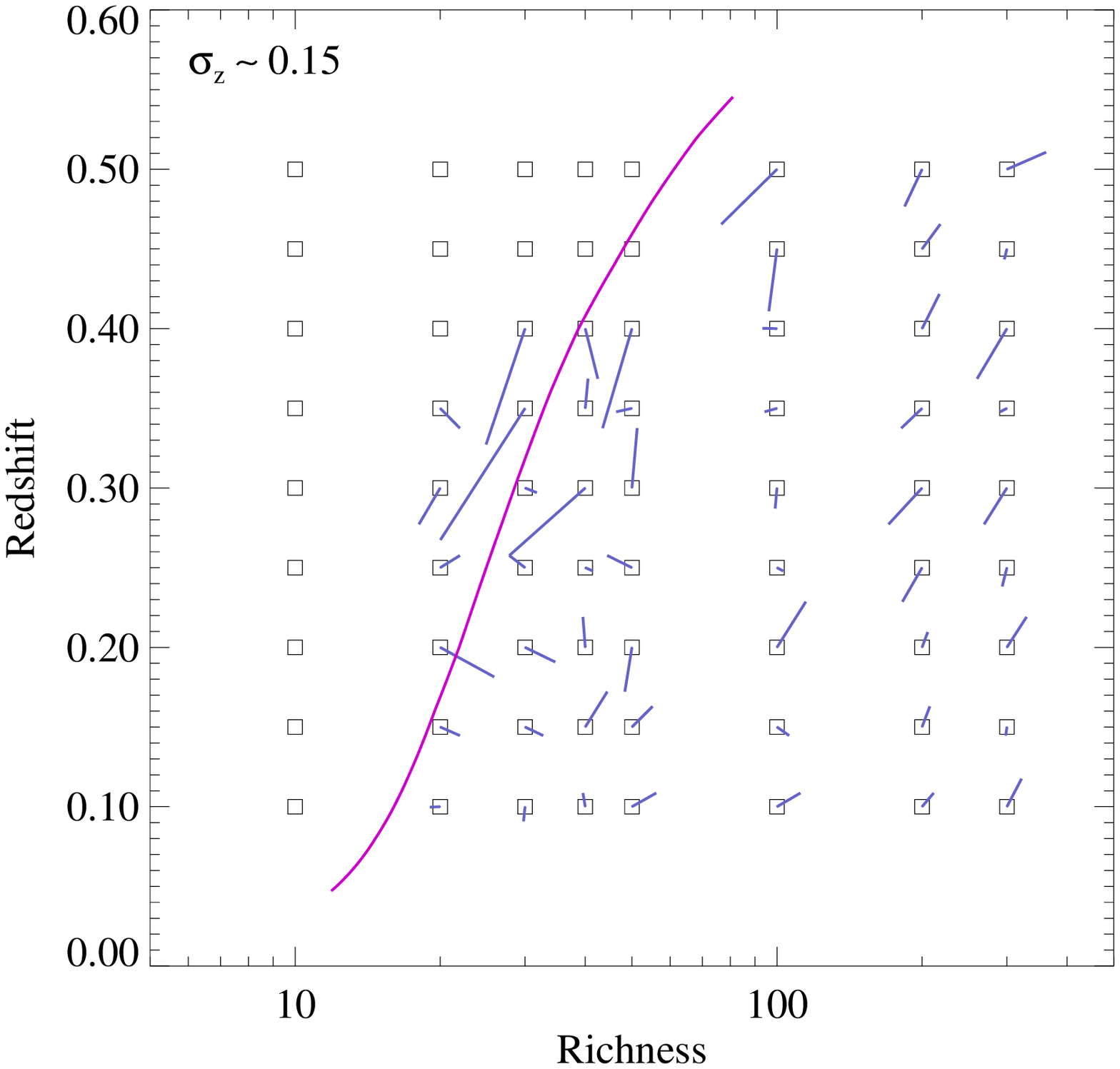}
\caption{
\label{fig:outputclusters:c}
  Richness and redshift of each of the input clusters (boxes) with the
short lines indicating the corresponding values determined from the AMF
fine filter. The long curved line indicates the approximate detection
limit and is computed by inserting $\sL_\cut = 100$ into equation (12)
and solving for $\Lc$ as a function $z_c$.
} \end{figure}

\begin{figure}
\plotone{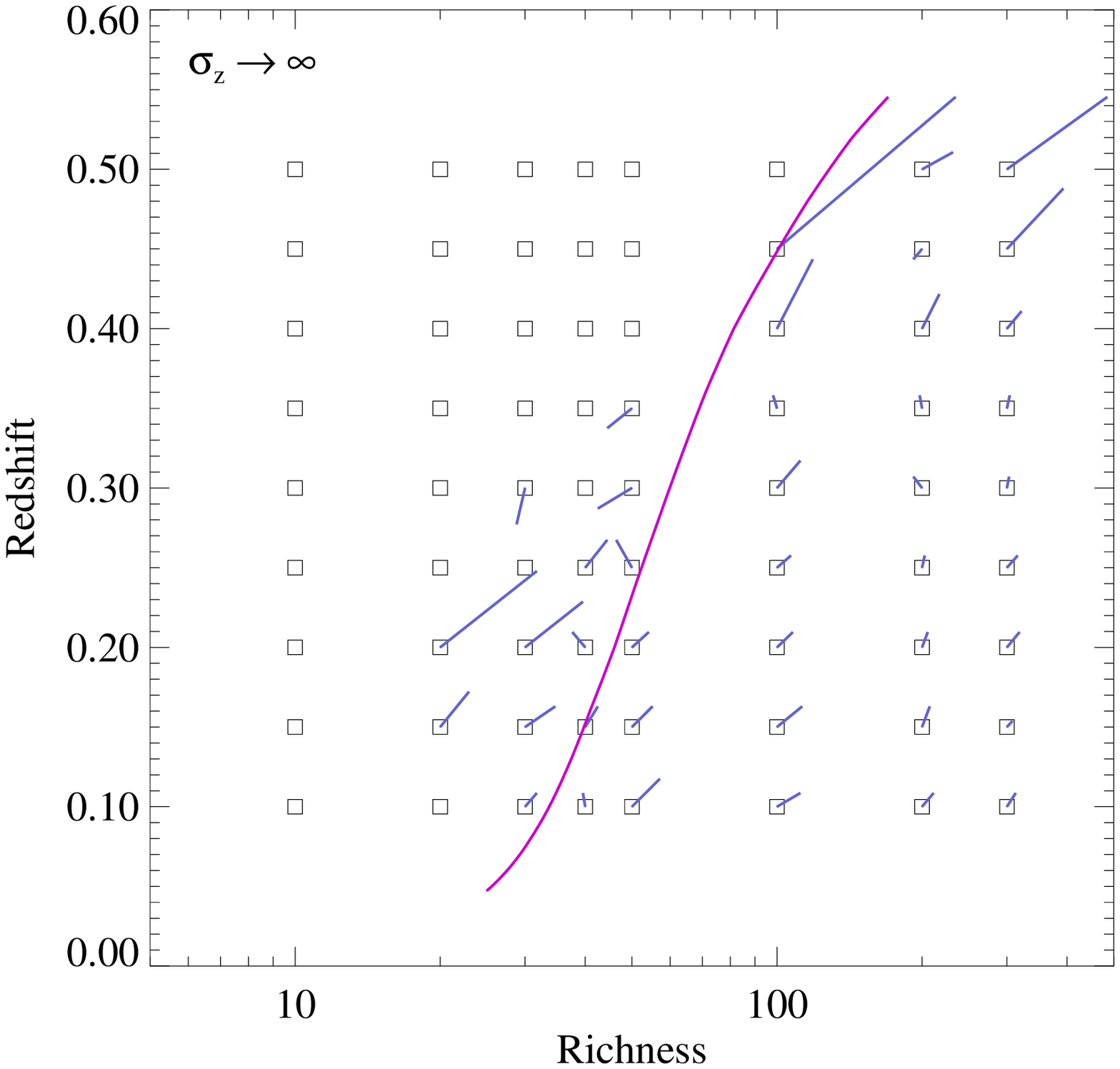}
\caption{
\label{fig:outputclusters:d}
  Richness and redshift of each of the input clusters (boxes) with the
short lines indicating the corresponding values determined from the
fine filter. The long curved line indicates the approximate detection
limit and is computed by inserting $\sL_\cut = 300$ into equation (12)
and solving for $\Lc$ as a function $z_c$.
} \end{figure}

  To test our AMF implementation we ran it on the above simulated galaxy
catalog for three different error regimes: $\bar{\sigma}_z \sim 0.05$,
$\bar{\sigma}_z \sim 0.15$ and $\bar{\sigma}_z \rightarrow \infty$. A
contour plot computed from the maximum likelihood points $\sL_\coarse^i$
is shown in figure~\ref{fig:outputclusters:i}, which indicates that the
coarse filter does a good job of finding the angular positions of the
clusters with no false detections.  The resulting cluster centers have
an accuracy that is within one core radius of the true cluster center. 
The redshifts and richnesses obtained by applying $\sL_\fine$ to the
clusters found with $\sL_\coarse$ are shown in
figures~\ref{fig:outputclusters:b}, \ref{fig:outputclusters:c}, and
\ref{fig:outputclusters:d}. The boxes denote the true redshift and
richness values of the input clusters.  The short lines connect the
inputs with the outputs (i.e. the values obtained with $\sL_\fine$).
Smaller lines indicate more accurate redshift and richness
determinations.  The long curve indicates the expected detection limit
for the value of $\sL_\cut$ used.  As expected, the number of clusters
detected and their accuracy decrease as $\bar{\sigma}_z$ increases. 
However, nearly all the clusters with $\Lc \gtrsim 100$ (corresponding
roughly to Abell richness class $\gtrsim 1$) are detected out to
redshifts of 0.5 which is the effective limit of the survey.  In a
deeper survey it will be possible to detect clusters at higher redshifts.

\begin{figure}
\plotone{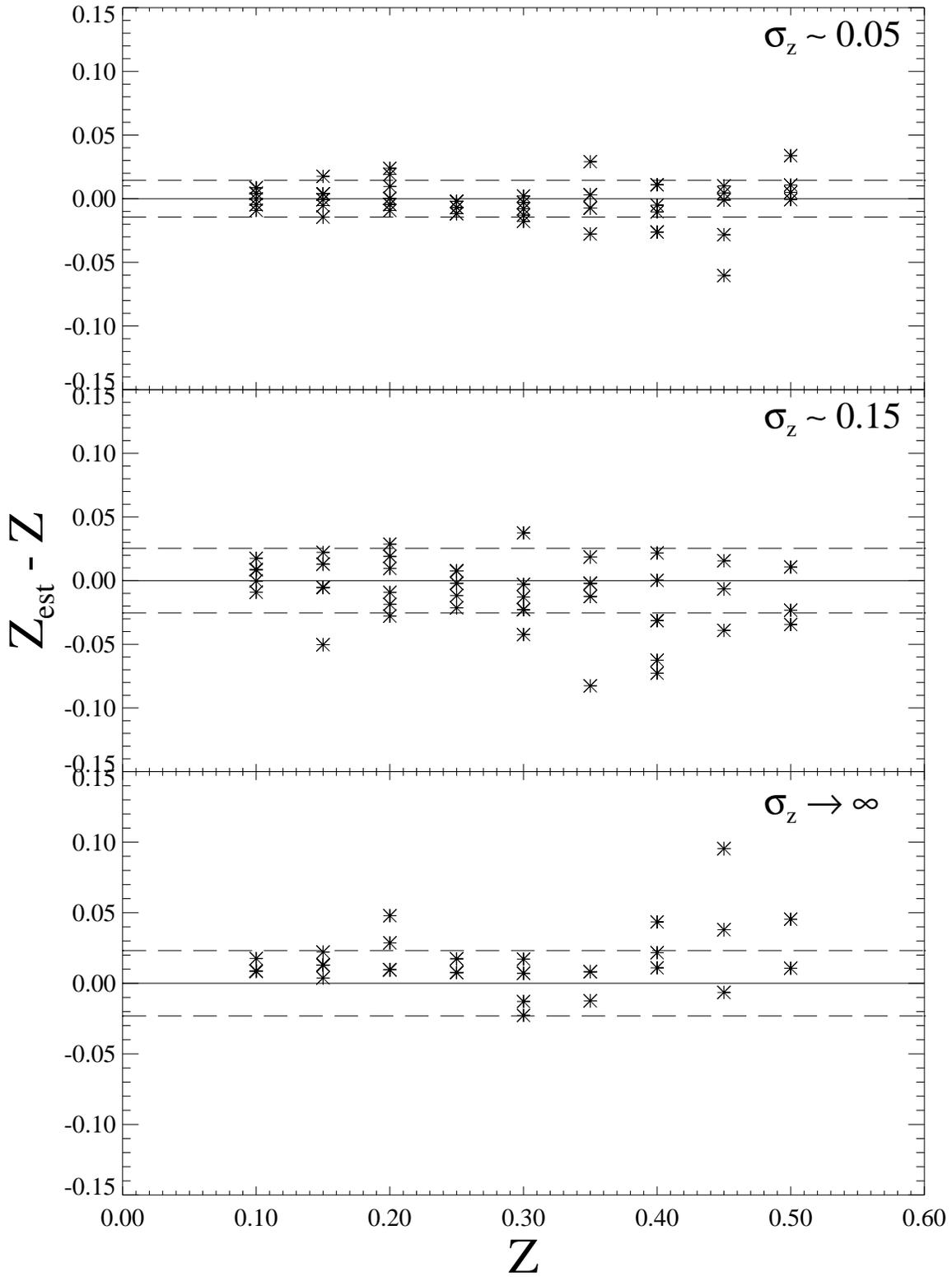}
\caption{
\label{fig:outputclusters:f}
  Redshift error as a function of redshift for all the detected
clusters.  Dashed lines indicate the 1-$\sigma$ error range computed from
these data, which from top to bottom are 0.014, 0.025 and 0.023.
} \end{figure}

\begin{figure}
\plotone{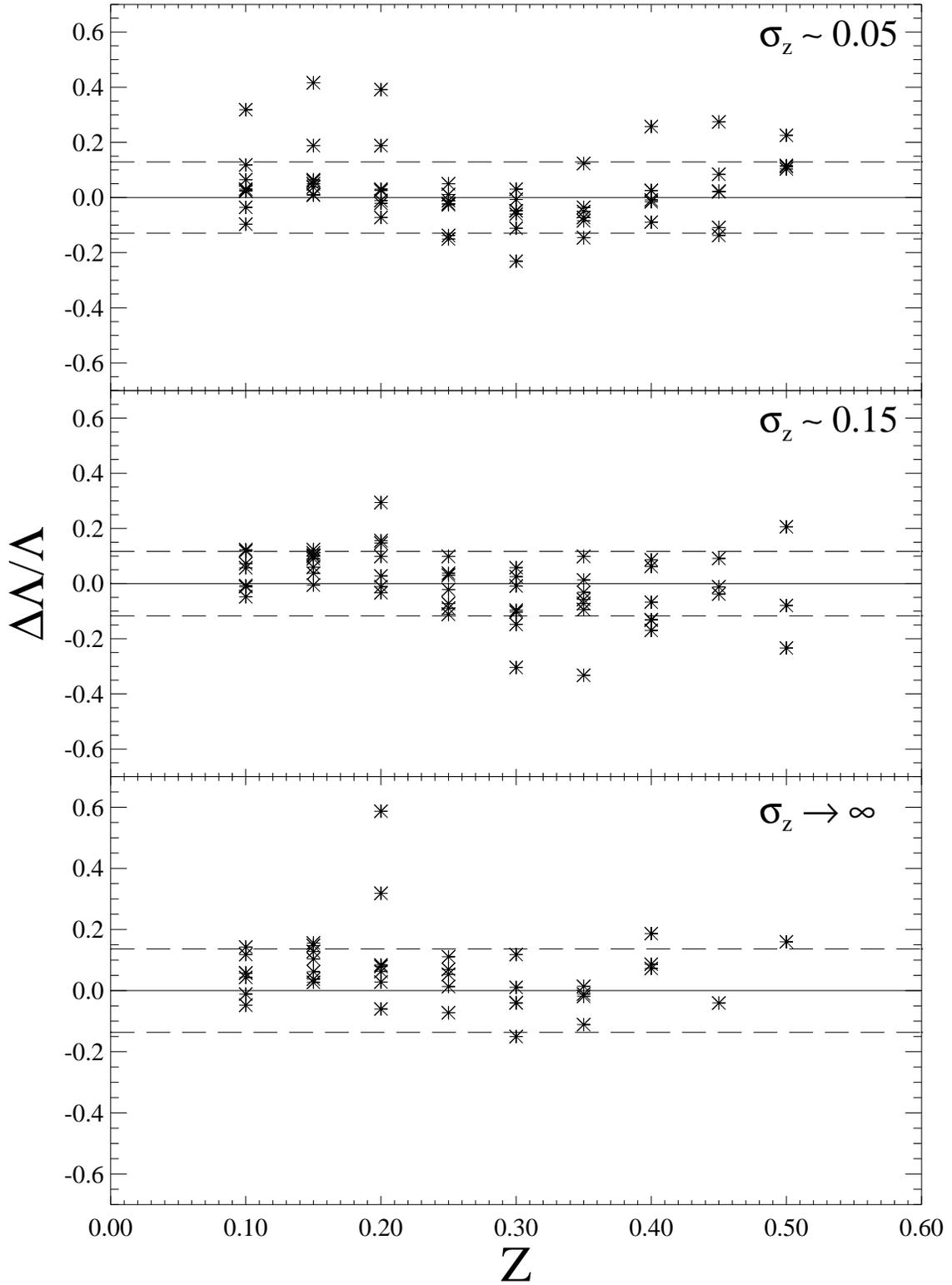}
\caption{
\label{fig:outputclusters:g}
  Fractional richness error as a function of redshift for all the
detected clusters.  Dashed lines indicate the 1-$\sigma$ range error
computed from these data, which from top to bottom are 0.13, 0.12 and
0.14.
} \end{figure}

  The errors in the redshift and richness estimates of all detected
clusters is presented in figures~\ref{fig:outputclusters:f} and
\ref{fig:outputclusters:g}.  A summary of these results is shown in
table~1.  In general, the AMF finds clusters with an accuracy of $\Delta
z \sim 0.02$ in redshift and $\sim$10\% in richness.  Including distance
information lowers the background and results in a substantial
improvement in the detection of poorer clusters.  Thus, many more
clusters are detected when $\sigma_z \sim 0.05$ as compared to $\sigma_z
\rightarrow \infty$.  These additional clusters are all poorer and thus
have higher errors, which explains why the average errors do not
significantly change with $\bar{\sigma}_z$. 

\begin{table}[tbh]
\begin{center}
Table 1: Summary of AMF tests on simulated data \\
~\\
\begin{tabular}{cccccc}
\hline
\hline
Input & Likelihood  &  Output    &  Output \\
error &  cutoff     & $z$ error &  $\Lambda$ error \\
\hline
$\bar{\sigma}_z$ & $\sL_\cut$ & $\sigma(\Delta z)$ & $\sigma(\Delta \Lambda/\Lambda)$ \\
\hline
 0.05    &  40 & 0.014 & 0.13 \\
 0.15    & 100 & 0.025 & 0.12 \\
$\infty$ & 300 & 0.023 & 0.14 \\
\hline
\hline
\end{tabular}
\end{center}
\end{table}

  Six additional tests were conducted on the $\bar{\sigma}_z \sim 0.05$
case to check the robustness of the results.  The first test explored
the effect of spatial structure in the background distribution of
galaxies by using positions taken from an N-body simulation (instead of
using randomly distributed field galaxies).  The second test
investigated the effect of using different K-corrections.  The next four
tests explored the effect of changing different parameters of the
cluster model: $\alpha$ in the Schechter luminosity function, $n$ in the
Plummer cluster density profile, the core radius ($r_\core$) and the
maximum radius of the cluster profile ($r_\max$).  The results of all
these tests are summarized in Table 2. 

  The N-body simulation contained $128^3$ dark matter particles in a
200~$\hMpc$ box (\cite{Xu95}) with sufficient spatial resolution to
resolve cluster cores.  The final $z = 0$ ouput of the simulation was
``stacked'' in a non-repeating fashion (\cite{Gott97}) to create a
simulated field out to $z = 0.6$.  Each particle was then assigned a
luminosity in the same manner as described earlier for the uniform
background.  The 72 test clusters were placed in the N-body background. 
The coarse filter was run on these data using the same parameters as in
the uniform case.  All the clusters detected with the uniform background
were also detected with the N-body background.  Next, the fine filter
was run on these clusters.  There was no change in the estimated
redshifts of the clusters.  The richness estimates showed a slight
increase ($\sim$ 20\%) in their variance, which is due to the
fluctuations in the background density. 


  In a real survey, it is unlikely that all the galaxies will have
correctly assigned Hubble types.  An error in the Hubble type primarily
affects the K-correction.  To test the effects of incorrect Hubble types
we ran the fine filter with an assumption that all the galaxies were
Ellipticals (E) and then again assuming that all the galaxies were
Spirals (Sc).  These changes produced no significant change in the
estimated redshift or richness of the clusters out to $z \sim 0.5$.

  In the real Universe, clusters can not be described by a single set of
model parameters.  Two tests of each of four model parameters were
carried out to look at the errors produced by using an AMF cluster model
with parameters different from the clusters in the data.  In each case
the fine filter was run on data with $\bar{\sigma}_z \sim 0.05$ and the
differences in the estimated redshift and richness were examined.  The
results are summarized in Table 2.  None of the changes in any of the
parameters significantly affected the estimated redshifts of the
clusters.  The largest effect on the richness $\Lambda$ occurred with
changing the parameter $\alpha$ in the Schechter luminosity function of
the cluster.  This induced a bias in $\Lambda$ at small redshift; the
bias decreases with redshift because the galaxies at high redshift are
not part of the faint-end Schechter luminosity function.  Changing the
slope of the cluster density profile had only a small effect on
$\Lambda$.  Changing the core radius had no significant effect on
$\Lambda$.  As expected, $\Lambda$ increases with increasing $r_\max$.

\begin{table}[tbh]
\begin{quote}
Table 2: Robustness tests on simulated data with $\sigma_z \sim 0.05$. 
A ``no change'' entry means that any difference in the estimated
redshift or richness was within the nominal errors quoted in Table 1
(i.e., $\Delta z \sim 0.02$ and $\Delta \Lambda/\Lambda \sim 0.1$). 
Biases are given relative to the nominal value (e.g., a bias of
1.1$\Lambda$ implies that the new estimated richnesss is 1.1 times the
nominal estimated richness).  All biases are independent of redshift and
richness unless stated otherwise. 
\end{quote}
\begin{center}
\begin{tabular}{llll}
\hline
\hline
Model Parameter   &   New  &     Effect on      &       Effect on \\
(nominal value)   &  Value & Estimated $z$ &  Estimated $\Lambda$ \\
\hline
\hline
 Background distribution & N-body         & no change & slight ($\sim$ 20\%) increase in variance \\
 (uniform)               &                &  & \\
\hline
 K correction            & all E          & no change & no change \\
 (E, Sa and Sc)          & all Sc         & no change & no change \\
\hline
 Cluster lum func slope  & -0.8           & no change &
    1.5$\Lambda$ bias at $z \sim 0.1$, \\
 ($\alpha = -1.1$)     &&& decreasing to no bias at $z \sim 0.35$ \\
                         & -1.3           & no change &
    0.5$\Lambda$ bias at $z \sim 0.1$, \\
&&& increasing to 1.2$\Lambda$ at $z \sim 0.5$ \\
\hline
 Cluster profile slope   &  1.5          & no change &
    1.1$\Lambda$ bias independent of $z$ \\
 ($n = 2.0$)             &  2.5          & no change &
    0.95$\Lambda$ bias independent of $z$ \\
\hline
 Cluster core radius     &  0.05 $\hMpc$  & no change & no change \\
 ($r_\core = 0.1~\hMpc$) &  0.20 $\hMpc$  & no change & no change \\
\hline
 Cluster max radius      &  0.75 $\hMpc$  & no change &
    0.75$\Lambda$ bias independent of $z$ \\
 ($r_\max = 1.0~\hMpc$)  &  1.25 $\hMpc$  & no change &
    1.25$\Lambda$ bias independent of $z$ \\
\hline
\hline
\end{tabular}
\end{center}
\end{table}

  The CPU and memory requirements of the AMF are dominated by the
$\sL_\coarse$ evaluation.  The AMF required around 100 MBytes of memory
and took from a few minutes ($\bar{\sigma}_z \sim 0.05$) to a little
under two hours ($\bar{\sigma}_z \rightarrow \infty$) using one CPU
(198Mhz MIPS R10000) of an SGI Origin200. For example, the SDSS will be
composed of approximately 1000 fields similar in size to our test
catalog.  Since finding clusters in one field is independent of all the
others, it is simple to run the AMF on a massively parallel computer;
it will be possible to run the AMF on the entire SDSS catalog in one
day.

\section{Summary and Conclusions}

  We have presented the Adaptive Matched Filter method for the automatic
selection of clusters of galaxies in a wide variety of galaxy catalogs.
The AMF can find clusters in most types of galaxy surveys: from
two-dimensional (2D) imaging surveys, to multi-band imaging surveys with
photometric redshifts of any accuracy (2$\half$D), to three-dimensional
(3D) redshift surveys.  The method can also be utilized in the selection
of clusters in cosmological N-body simulations. The AMF is based on
matching the galaxy catalog with a cluster filter that models the
overall galaxy distribution. The model describes the surface density,
apparent magnitude, and redshift of cluster and field galaxies. 
Convolving the data with the filter produces a cluster probability map
whose peaks correspond to the location of the clusters. The probability
peaks also yield the best fit redshift and richness of each cluster.

  The heart of the AMF is the apparent overdensity $\delta_i$ which is
evaluated at each galaxy position and has a higher value for galaxies
in clusters than galaxies in the field.   The apparent overdensity
distills the entire description of the galaxy catalog into a single
function. Two likelihood functions are derived, $\sL_\coarse$ and
$\sL_\fine$, using different underlying model assumptions. The
theoretical framework of the AMF allows  estimated redshifts to be
included via a simple redshift filter, which effectively limits the
sums in $\sL_\coarse$ and $\sL_\fine$ to those galaxies within a window
around $z_c$.  The maxima in the likelihood functions are used to identify
cluster positions as well as their redshifts and richnesses.

  The AMF is adaptive in three ways.  First, it adapts to the errors
in the estimated redshifts.  Second, it uses the locations of the
galaxies as ``naturally'' adaptive grid to ensure sufficient resolution
at even the highest redshifts.  Third, it uses a two step approach that
applies a coarse filter to initially find the clusters and a fine filter
to more precisely estimate the redshift and richness of each cluster.

  We tested the AMF on a set of simulated clusters with different
richnesses and redshifts---ranging from groups to rich clusters at
redshifts 0.1 to 0.5; the clusters were placed in a simulated field of
randomly distributed galaxies as well as in a non-random distribution
produced by N-body cosmological simulations.  We find that the AMF
detects clusters with an accuracy of $\Delta z \sim$0.02 in redshift and
$\sim$10\% in richness to $z \lesssim 0.5$ (for a simulated galaxy
survey to $r' \approx 23.5$).  In addition, robustness tests provide a
strong indication that the AMF will perform well on observational data
sets.  Detecting clusters at even higher redshifts will be possible in
deeper surveys. 

\acknowledgments

  We wish to thank Guohong Xu for providing the N-body simulations used
in this paper.  In addition David Weinberg, Andy Connolly, Marc Postman,
Lori Lubin, and Chris Finger provided helpful suggestions.  We would
also like to thank Dr.  Micheal Kurtz for his helpful comments.  This
work was supported in part by the DoE Computational Science Fellowship
Program, the Princeton Observatory Advisory Council, and NSF grants
AST~93-15368 and GER~93-54937. 

\appendix

\section{Cluster Profile}
The projected cluster profile is derived from spherical
modified Plummer law profile:
  \BEQ
        \rho_c(r) = \left\{
                 \begin{array}{c@{\quad,\quad}c}
                   \frac{\rho_c^0}{(1 + r^2/r_\core^2)^\frac{n}{2}}
                 - \frac{\rho_c^0}{(1 + r_\max^2/r_\core^2)^\frac{n}{2}} &
                     r \leq r_\max \\
                     0 & r > r_\max
                 \end{array}
                 \right.
  \EEQ
where $n \approx 2$ and the constant $\rho_c^0$ is used to normalize
the profile. The projected profile is computed by integrating along the
line of sight:
  \BEQ
        \Sigma_c(r) = 2 \int_{r}^{\infty}
                            \frac{\rho_c(t) t dt }
                             { \sqrt{t^2 - r^2} } ,
  \EEQ
which for the modified plummer profile gives
  \BEQ
        \Sigma_c(r) = \left\{
                 \begin{array}{c@{\quad,\quad}c}
                   \frac{\Sigma_c^0}{(1 + r^2/r_\core^2)^\frac{n-1}{2}}
                 - \frac{\Sigma_c^0}{(1 + r_\max^2/r_\core^2)^\frac{n-1}{2}} &
                     r \leq r_\max \\
                     0 & r > r_\max
                 \end{array}
                 \right.
  \EEQ
where $\Sigma_c^0$ is a constants set by the normalization.


\section{Luminosity Distance}
  The transformation between absolute and apparent luminosities in band
$\lambda_0$
  \BEQ
       L_{\lambda_0}  = 4 \pi D^2(z;\mu) l_{\lambda_0},
  \EEQ
where $\mu$ is the galaxy type (e.g. elliptical, spiral, irregular, ...) and
$D$ is the luminosity distance at a redshift $z$ and is related to the 
angular size distance $d$ by
  \BEQ
        D^2(z;\mu) = \frac{(1+z)^2 d^2(z)}
                          {K_{\lambda_0}(z;\mu) E_{\lambda_0}(z;\mu)} ~ ,
  \EEQ
where $K_{\lambda_0}$ and $E_{\lambda_0}$ are the``K'' and evolutionary
corrections. For no evolution models $E_{\lambda_0} = 1$. The
angular-size distance for $\Omega + \Omega_\Lambda = 1$ cosmologies is
given by
  \BEQ
       d(z) = \frac{c}{H_0} \int_0^z \frac{dz'}
                   {(1 - \Omega + \Omega (1+z')^3)^{1/2}}
  \EEQ




\section{Derivation of the Likelihood Functions}
  Various likelihood functions can be derived.  The differences are due
to the additional assumptions that are made about the distribution of
the observations.  This appendix gives the mathematical derivation of
the two likelihood functions used in the AMF: $\sL_\coarse$ and
$\sL_\fine$.  Both derivations are conceptually based on virtually binning the
data, but make different assumptions about the distribution of galaxies
in the virtual bins.

  Imagine dividing up the angular and apparent luminosity
domain around a cluster into bins.  We assign to each bin a unique
index $j$.  The expected number of galaxies in bin $j$
is denoted $n_\model^j$.  The number of galaxies actually found in bin $j$
is $n_\data^j$.  In general, the probability of finding $n_\data^j$
galaxies in cell $j$ is given by a Poisson distribution
  \BEQ
        P_j = \frac{(n_\model^j)^{n_\data^j} e^{-n_\model^j}}
                   {       n_\data^j ! }
  \EEQ
The likelihood of the data given the model is computed from the sum 
of the logs of the individual probabilities
  \BEQ
       \sL = \sum_j \ln P_j .
  \EEQ

\subsection{Coarse Grained $\sL$}
  If the virtual bins are made big enough that there are many galaxies
in each bin, then the probability distribution can be approximated by a
Gaussian
  \BEQ
        P_j = \frac{1}{\sqrt{2\pi n_\model^j}}
              \exp \left \{
                 - \frac{(n_\data^j - n_\model^j)^2}
                        { 2 n_\model^j}
                   \right \} ~ .
  \EEQ
Furthermore, if the field contribution is approximately
uniform and large enough to dominate the noise then
  \BEQ
        P_j = \frac{1}{\sqrt{2\pi n_f^j}}
              \exp \left \{
                   - \frac{(n_\data^j - n_\model^j)^2}{ 2 n_f^j}
                  \right \} ~ .
  \EEQ
where $n_\model^j = n_f^j + \Lc n_c^j$. Summing the logs of these
probabilities results in the following expression for the coarse
likelihood
  \BEQA
       \sL_\coarse
       & = & \sum_j \ln P_j \NN \\
       & = & -\frac{1}{2} \sum_j \ln 2 \pi n_f^j
         -   \frac{1}{2} \sum_j 
               \frac{(n_\data^j - n_\model^j)^2}{ n_f^j}
  \EEQA
The first term does not affect the value of $\Lc$ which minimizes
$\sL_\coarse$ and can be dropped. In addition, if the bins can also be made
sufficiently small, then the sum over all the bins can be replaced by an
integral
  \BEQ
        \sL_\coarse = -\int
            \frac{(n_\data(\theta,\l) - n_\model(\theta,l))^2}
                 { n_f(\theta,l)} d\Omega dl ~ ,
  \EEQ
where $n_\data(\theta,l)$ is a sum of Dirac delta functions
corresponding to the locations of the galaxies. Expanding the squared
term and replacing $n_\model$ with $n_f + \Lc n_c$ yields
  \BEQ
        \sL_\coarse = -\int
            \frac{n_\data^2 - 2 n_\data n_f - 2 n_\data \Lc n_c
                  + n_f^2 + 2 n_f \Lc n_c + \Lc^2 n_c^2}
                 { n_f} d\Omega dl ~ .
  \EEQ
The above expression can be simplified by setting $\delta = n_c/n_f$,
dropping all expressions that are independent of $\Lc$, and noting that
$\Lc \int n_c$ is small compared to the other terms, which leaves
  \BEQ
        \sL_\coarse = 2 \Lc \sum_i \delta_i
                    - \Lc^2 \int  \delta(\theta,l) n_c(\theta,l) d\Omega dl
  \EEQ
Taking the derivative of $\sL_\coarse$ with respect to $\Lc$ and
setting the result equal to zero, we can solve for $\Lambda_\coarse$
directly
  \BEQ
        \Lambda_\coarse = \frac{
              \sum_i \delta_i
           }{
              \int \delta(\theta,l) n_c(\theta,l) d\Omega dl
          } .
  \EEQ
where the denominator terms of $\phi_c$, $\Sigma_c$ and $n_f$ is
  \BEQA
         \int \delta(\theta,l) n_c(\theta,l) d\Omega dl
               &=& \int \frac{n_c^2}{ n_f} d\Omega dl \NN \\
               &=&
          \left [
             \int^\infty_{L(l_{min},z_c)} 
             \frac{\phi_c^2(L')}{n_f(L')} dL'
          \right ]
          \left [ 
             \frac{ d_c^2}{(1 + z_c)^2}
             \int^{r_\max}_0 \Sigma_c^2(r') 2 \pi r' dr'
          \right ] ~ .
  \EEQA
Finally, inserting the above value of $\Lambda_\coarse$ back into the
expression for $\sL_\coarse$ results in
  \BEQ
        \sL_\coarse = \Lc \sum_i \delta_i ~~,
  \EEQ
which is $\Delta_\data$.  Thus, maximizing the measured cluster
overdensity will give the correct richness and location of the cluster.

\subsection{Fine Grained $\sL$}

  If the virtual bins are chosen to be sufficiently small that no bin
contains more that one galaxy, then the calculation of $\sL$ can be
significantly simplified because there are only two probabilities that
need to be computed. The probability of the empty bins
  \BEQ
        P_\none  =  e^{-n_\model^j} 
  \EEQ
and the probability of the filled bins
  \BEQ
        P_\filled  = n_\model^j e^{-n_\model^j} .
  \EEQ
The sum of the log of the probabilities is then
  \BEQA 
        \sL_\fine &=& \sum_\none \ln P_\none
                   +  \sum_\filled \ln P_\filled \NN \\
                  &=& - \sum_\none n_\model^j
                      - \sum_\filled n_\model^j
                      + \sum_\filled \ln n_\model^j
  \label{eq:finesum}
  \EEQA

  By definition summing over all the empty bins and all the filled bins
is the same as summing over all the bins.  Thus, the first two terms in
equation (\ref{eq:finesum}) are just the total number of galaxies
predicted by the model
  \BEQA
        \sum_{\rm all~bins} n_\model^j
        &=& \int^\infty_{l_\min} \int^{\theta_\max}_0 n_\model(\theta',l';z_c) d\Omega' dl' \NN \\
        &=& N_f + \Lc N_c
  \EEQA
where $l_{min}$ is the apparent luminosity limit of the survey.
$N_f$ and $\Lc N_c$ are the total number of field and cluster galaxies one expects
to see inside $\theta_\max$; they can be computed by integrating $n_f$ and 
$n_c$
  \BEQA
       N_f(z_c) &=& \pi \theta_\max^2(z_c) \int^\infty_{l_\min} n_f(l') dl' ~ , \NN \\
       N_c(z_c)
       &=& \int^\infty_{l_\min} \int^{\theta_\max}_0 n_c(\theta',l';z_c) d\Omega' dl' \NN \\
       &=& \Phi_c(L(l_\min,z_c)) \int^{r_\max}_0 \Sigma_c(r') 2 \pi r' dr'
  \EEQA
where $\Phi_c(L) = \int^\infty_L \phi_c(L') dL'$.

  Because we retain complete freedom to locate the bins wherever we like, we
can center all the filled bins on the galaxies, in which case the third
term in equation (\ref{eq:finesum}) becomes
  \BEQ
        \sum_\filled \ln n_\model^j \rightarrow \sum_i \ln[n_f^i + \Lc n_c^i] ~ ,
  \EEQ
and the sum is now carried out over all the galaxies instead of all the
filled bins. Combining these results we can now write the likelihood in
terms that are readily computable from the model and the galaxy catalog
  \BEQ
        \sL_\fine = - N_f - \Lc N_c + \sum_i \ln[n_f^i + \Lc n_c^i] ~ .
  \EEQ

  The simplest way to find the richness is to take the derivative of the
above equation with respect to $\Lc$ and setting it equal to zero yields
  \BEQ
       N_c = \sum_i \frac{\delta_i}{1 + \Lambda_\fine \delta_i} ~ ,
  \EEQ
where we have substituted $\delta_i$ for $n_c^i/n_f^i$.  Unfortunately,
it is not possible to solve for $\Lambda_\fine$ explicitly, but a numerical
solution can be found by standard methods.  The resulting value of
$\Lambda_\fine$ is then inserted into the following equation for $\sL_\fine$
to obtain the maximized value of the likelihood
  \BEQ
        \sL_\fine = - \Lambda_\fine  N_c + \sum_i \ln[1 + \Lambda_\fine \delta_i] ~ .
  \EEQ
[Note: in the above expression terms that are independent of $\Lambda_\fine$
and do not contribute any additional information to $\sL_\fine$ have been
dropped.]

  Finally, it is worth mentioning again that while $\Lc_\coarse$ can be
obtained directly from $\sL_\coarse$, $\Lc_\fine$ can only be found by
numerically finding the zero point of equation (C19).  Furthermore,
equation (C19) does not lend itself to standard derivative based solvers
(e.g., Newton-Raphson) that produce accurate solutions in only a few
iterations.  Fortunately, the solution can usually be bracketed in the
range $0 < \Lc_\fine < 1000$, thus obtaining a solution with an accuracy
$\Delta \Lc \sim 1$ takes $\log_2 (1000/1) = 10$ iterations using a
bisection method.

\section{Data/Model Consistency Checks}
  The first consistency check can be made with $n_f(l)$, or its
cumulative probability $N_f(l) = \int_l^\infty n_f(l') dl'$, which can
be fit directly to the galaxy catalog.  In the case where a simulated
catalog is used the luminosity distribution is taken from the
underlying field luminosity function $\phi_f(L)$ and it is possible to
compute $N_f(l)$ for each Hubble type directly
  \BEQ
        N_f(l) = \int_0^\infty  \Phi_f( L(l,z,\mu) ) d(z)^2 \frac{d(z)}{dz} dz ~ ,
  \EEQ
where $\Phi_f(l) = \int_l^\infty \phi_f(l') dl'$.

  The second consistency check can be made with the distribution
in redshift (if estimated redshifts exist), where the number of galaxies
at each redshift should satisfy the following convolution with the
mean estimated error
  \BEQ
        \frac{dN_f}{dz}(z) = \int_{-\infty}^\infty
           \frac{1}{\sqrt{2\pi} \bar{\sigma}_z}
              \exp \left \{
                 - \frac{(z - z')^2}
                        { 2 \bar{\sigma}_z^2 }
                   \right \}
           \Phi_f( L(l_\min,z',\mu) )  \frac{d(z')}{dz} dz' ~ .
  \EEQ



\newpage


\end{document}